\documentclass[aps,prb,twocolumn,superscriptaddress,showpacs,floatfix]{revtex4-1}

\usepackage{graphicx,graphics,color,epsfig}
\usepackage{color}
\usepackage{times}
\usepackage{bm}
\usepackage{amsmath}

\usepackage{amssymb}
\usepackage{epstopdf}

\begin{document}

\title{The Electronic Correlation Strength of Pu}

\author{A. Svane}
\email[Electronic address:]{svane@phys.au.dk}
\affiliation{Department of Physics and Astronomy, Aarhus University, DK 8000 Aarhus C, Denmark}

\author{R. C. Albers}
\affiliation{Theoretical Division, Los Alamos National Laboratory, Los Alamos, NM 87545 USA}

\author{N. E. Christensen}
\affiliation{Department of Physics and Astronomy, Aarhus University, DK 8000 Aarhus C, Denmark}

\author{M. van Schilfgaarde}
\affiliation{Department of Physics, King's College London, The Strand, London WC2R 2LS, UK        }

\author{A. N. Chantis}
\affiliation{American Physical Society, 1 Research Road, Ridge, New York 11961, USA}

\author{Jian-Xin Zhu}
\affiliation{Theoretical Division, Los Alamos National Laboratory, Los Alamos, NM 87545 USA}

\date{\today}

\begin{abstract}
An electronic quantity, the  correlation strength, is defined as a necessary
step for understanding the properties and trends in strongly correlated electronic materials. 
As a test case, this is applied to the different phases of elemental Pu.
Within the $GW$ approximation we have surprisingly found a ``universal" scaling relationship, 
where the {\it f}-electron bandwidth reduction due to correlation effects 
is shown to depend only on the local density approximation bandwidth and 
is otherwise independent of crystal structure and lattice constant.
\end{abstract}


\maketitle

\section{Introduction}

Many technologically important materials have strong electron-electron correlation effects.  
They exhibit large anomalies in their physical properties when compared with 
materials that are weakly correlated, and
have significant deviations in their electronic-structure from that predicted by conventional band-structure theory  
based on the local-density approximation (LDA). 
Because the anomalies and deviations are caused by electronic correlation effects,
which often dominate the physics of these materials,
in this paper we define a quantity that we call the ``correlation strength,"
or $C$, as a necessary step in order to be able to describe trends 
and bring order into our understanding of correlated materials.
We emphasize the word ``quantity" since a quantitative measure is needed
to answer the question, ``How strong are the electronic correlations?" 
Without some understanding of how big this is, it is not possible
to make sense of the properties of these materials.
In this context, ``correlation'' is defined in a way somewhat different from how it
is sometimes used (e.g., in the term ``exchange-correlation potential").  
By ``correlation" we specifically mean ``correlation beyond LDA theory."
This usage reflects the way the term is often loosely used in common terminology in the area of strongly
correlated electronic systems.

To create a new quantity requires determining a ``scale" by which to measure its size.  
In principle, any experimental or theoretical property (e.g., specific heat) that monotonically increases or decreases
over the full range of correlation effects, where we define correlation strength to lie between zero for none and one for full correlation,
can be used as a measure of this quantity.  Hence correlation strength is an indeterminant quantity
and depends on the property used to define it.  
However, this does not matter since only relative rather than any absolute strength is important 
for characterizing these materials and for predicting trends in their properties. 
Any measure based on one property can easily be converted to that based on another property.
In this paper we develop a theoretical correlation strength based on the $GW$ approximation \cite{hedin65,hedin69,hedin99,aryasetiwan98}
to electronic-structure theory and apply it to 
plutonium, \cite{chantis09,kutepov12} which is known to have significant correlation effects. 
The $GW$ approximation is named for the correction term in this theory, 
which is a Green's function \textit{G} times a screened Coulomb interaction \textit{W}.
We also demonstrate a 
scaling relationship that is universal in that it is independent of crystal structure and atomic volume.
The ideas in this paper could certainly be modified and generalized to be able to treat other types of correlated
materials (e.g., spin-fluctuation or high-temperature superconducting materials) by using other electronic
properties to determine a correlation strength and by using
more sophisticated theoretical techniques than are considered here.

Of course, there is a long history in physics and chemistry of using various quantities
to predict materials trends.  For example, with respect to the actinides, in 1970
 Hill\cite{hill70} plotted the magnetic and superconducting transition temperatures of actinide compounds 
as a function 
of the actinide-actinide nearest-neighbor distance.  These ``Hill plots" brought some sensible order into what
had previously been seen as a somewhat random occurrence of these various ground states, and also
provided some degree of predictability, in that superconducting compounds tended to occur for short actinide spacings
and magnetic compounds at large spacings.\cite{boring2000}  The plots were intuitively based on the
idea that $f$-wave-function overlap was the key factor determining the stability of the relative ground states.
These plots failed for heavy-fermion compounds\cite{boring2000} and our understanding of 
electronic structure has now advanced to the point where we realize that at large actinide nearest-neighbor
distances the $f$ electrons tend to hop predominantly through hybridizations with other orbitals on nearby atoms
rather than through a direct $f$-$f$ hybridization.

Another important actinide trend was developed by Smith and Kmetko.\cite{smith83} 
They showed that the crystal structures of the actinides can be plotted as a continuous function of atomic
number ($Z$), with alloys filling in between the atomic numbers of the pure elements.
When plotted in this way, one obtains ``connected binary alloy phase diagrams for the light actinides," 
which provide a clear picture of the trends and relationships
between the crystal structures of all the light actinides ``at a glance."

More generally, in materials science, many different variables have been used in an attempt to
understand systematic trends in crystal structures among classes of different compounds.
Such variables have included electronegativity differences, covalent and ionic contribution to the average
spectroscopic energy gap, and various types of core, ionic, and metallic radii. These have been
reviewed in a review article on ``Structure Mapping" by Pettifor;\cite{pettifor00}
see also Refs.~\onlinecite{villars83,pettifor86,christensen87,pettifor88,fischer06}.
However, these methods are not relevant for our purposes, since, as we shall show below, 
correlation effects are more important than crystal structure for determining the properties of
many actinide metals.

Among different classes of correlated materials, superconducting transition pressures
have often been plotted versus either specific structural properties or some characteristic
correlated quantity. These are too numerous to report in full.  A typical example are trends
in  superconducting transition temperatures\cite{fischer07,takahashi08} 
with numbers of planar (layered or two-dimensional)
structural units (e.g., CuO$_2$ or FeAs planes), and similarly for representative classes of
some heavy-fermion superconductors (e.g., CeMIn5 and PuMGa5 for M=Co, Rh, Ir,  
also including c/a structural anisotropies\cite{bauer04}). 
Closer in spirit to this paper are trends in superconducting transition temperature versus characteristic
spin-fluctuation energies, except that the trends were all based on experimental
measurements rather than theoretical input.\cite{bauer04,sarrao07,pfleiderer09}

Perhaps the closest analog to the ideas of our paper is the correlation between crystal structure
and $d$-occupation numbers in rare-earth systems (including under 
pressure).\cite{duthie77,skriver85}
In this case theoretical calculations are required to determine the number of occupied $d$ electrons
as a function of $d$ element and volume per atom (which can be equated to pressure).  Given
this input, however, the correct crystal structure can then usually be predicted. 
What is different about our approach is that we believe that not just one property such as crystal structure or
transition temperature, but many properties of actinide metals will follow trends based on our correlation scale (see below).

The outline of the paper is as follows: In Sec.~\ref{sec:method}, a theoretical definition of the correlation scale 
is presented. It is expressed in terms of the effective band width based on the parameter-free LDA and $GW$ approaches. 
In Sec.~\ref{sec:results}, we apply the scenario to determine the correlation strength in elemental Pu solids.  
A universal scaling relationship is obtained, where the $f$-electron bandwidth reduction 
due to correlation effects is shown to depend only upon the LDA bandwidth 
and is otherwise independent of crystal structure and lattice constant.  
The same type of trend is also found for the $d$-electron systems. 
A concluding summary is given in Sec.~\ref{sec:summary}. 

\section{Theoretical Method}
\label{sec:method}

Our meaning of correlation makes it necessary to use a theory that includes correlation effects
that go beyond those included by the LDA approximation in order
to determine a theoretical correlation strength.
This is challenging, since the most sophisticated treatments of correlation
effects have historically been mainly confined to abstract theoretical models, 
and have parameterized the electronic structure in such an oversimplified manner 
that the connection with actual materials examined experimentally 
was often somewhat vague.\cite{albers09} 
In the last decade, however, great progress has been made in this area, 
especially those involving
dynamical mean-field theory (DMFT)~\cite{georges96,kotliar06,held08,kunes10}
techniques, and strong correlation effects are beginning to be integrated into true first
principles methods.  To achieve this, instead of using {\it ad hoc} Hubbard Hamiltonians that
were essentially added without derivation to local density approximation
calculations, more recent methods have been attempting to explicitly
calculate screened Coulomb interactions directly in the random phase approximation
(RPA) and related approximations.  These techniques have been recently
reviewed by Imada and Miyake.\cite{imada10}
One direction that has been particularly fruitful recently is the construction of low-energy
effective models involving a downfolding of the electronic states and using
localized Wannier orbitals and {\it ab initio} real-space tight-binding models. States far
from the Fermi energy can be treated with conventional LDA-like techniques,
while correlation effects are taken explicitly into account for the important
states around the Fermi energy.  Usually constrained RPA (or cRPA) methods
are used to screen the Coulomb interactions.  Such methods have achieved a
fair degree of success for semiconductors, 3$d$ transition-metal oxides, iron-based
superconductors, and organic superconductors.  

However, these methods
rely upon being able to separate the electronic structure into some electrons belonging
to fairly isolated bands near the Fermi level and the rest to band degrees
of freedom far from the Fermi level.  For metals, as we are considering, such
methods therefore appear to be unlikely to be successful.  
Another approach,\cite{sun02,biermann03}
which seems more suitable to our case, is GW+DMFT. 
This has also been reviewed in Ref.~\onlinecite{imada10}.
Such a method involves $GW$ (or RPA-like) methods for calculating
the Coulomb interactions that are then integrated with DMFT 
techniques.  In the full implementation the entire scheme
would be made self-consistent and would be independent
of the initial $GW$ calculations used to initiate the method. 
In the initial description of the method\cite{biermann03}
only a simplified one-shot approach was applied to nickel.
Since the initial papers outlining the methodology,
almost no progress has been made, perhaps
indicating the difficulty of this approach.
Very recently, however, a more sophisticated implementation\cite{tomczak12}
has been applied to SrVO$_3$.  While these
calculations are not yet fully self-consistent,
they may stimulate more interest in pushing through the technical
issues involved in implementing this method.

Since there  is not yet widely available a suitable code
that involves these more sophisticated treatments
of correlation for the metallic systems 
that we are interested in, 
we have used the $GW$ method\cite{hedin65,aryasetiwan98,hedin99}
as a  theoretical method for estimating correlation effects.
Although this is a low-order approximation that definitely fails for very strong correlation effects,
it is sufficient for our purposes as a way to estimate correlation deviations from LDA band-structure theory,
and in particular for the main  
purpose of our work, which is to show that it is possible and useful to define a new quantity, 
which we call correlation strength,
in order to be able to place new materials in their proper physics context
and hence to be able to observe important trends in their properties.

Among the available $GW$ codes, we have used the quasiparticle self-consistent $GW$ approximation 
(QSGW).\cite{vanschilfgaarde06a,vanschilfgaarde06b,kotani07} 
The $GW$ approximation, itself, can be viewed as the first term in the expansion of the nonlocal
energy-dependent self-energy $\Sigma(\bf{r},\bf{r}^{\prime},\omega)$ in the screened Coulomb interaction $W$. 
From a more physical point of view it can
also be interpreted as a dynamically screened Hartree-Fock approximation 
plus a Coulomb hole contribution.\cite{hedin99,aryasetiwan98}  
Therefore, $GW$ is a well defined perturbation theory.  
In its usual implemention, sometimes called the ``one-shot" approximation, 
it depends on the one-electron Green's functions which use LDA eigenvalues and eigenfunctions, 
and hence the results can depend on this choice.  
Unfortunately, as correlations become stronger serious practical and formal problems 
can arise in this approximation.~\cite{vanschilfgaarde06b} 
However, Kotani {\em et al.}~\cite{kotani07} have provided a way to surmount this difficulty, 
by using a self-consistent one-electron Green's function
that is derived from the self-energy (the quasi-particle eigenvalues and eigenfunctions) 
instead of LDA as the starting point. 
In the literature, it has been demonstrated that the QSGW form of $GW$ theory 
reliably describes a wide range of semiconductors,~\cite{vanschilfgaarde06a,svane10,svane10b,svane11} 
 $spd$,~\cite{vanschilfgaarde06a,faleev04,chantis06} and rare-earth systems.~\cite{chantis07}
It should be noted that the energy eigenvalues of the QSGW method 
are the same as the quasiparticle spectra of the $GW$ method.  
This captures the many-body shifts in the quasiparticle energies.  
However, when presenting the quasiparticle DOS, this ignores the smearing 
by the imaginary part of the self-energy of the spectra due to quasiparticle lifetime effects, 
which should increase as quasiparticle energies become farther away from the Fermi energy.

To define a theoretical correlation strength  some electronic-structure quantity that scales with an intuitive
notion of correlation strength is needed. 
In our application to Pu, we propose to consider the $f$ bandwidth, $W_f$, 
and use the relative bandwidth reduction in QSGW compared to LDA, 
\begin{equation}
{\textrm w_{rel}}= W_f(\text{GW})/W_f(\text{LDA}), 
\end{equation}
as the key quantity, where $W_f(\text{GW})$ and $W_f(\text{LDA})$ 
are the $f$ bandwidths as obtained from QSGW and LDA calculations, respectively.
This is consistent with the correlation-induced QSGW $f$-bandwidth reduction in Pu 
that was demonstrated in Ref.~\onlinecite{chantis09}.

Using a quasiparticle calculation is important since lifetime effects,
which are absent in the LDA calculations, would obscure 
the band narrowing in $GW$ relative to LDA.
We also need a measure that is robust at the high temperatures
of the strongly correlated phases of Pu, where any low-energy
features in the electronic structure are likely to be thermally averaged
away.~\footnote{The most strongly correlated phase of Pu is $\delta$ Pu, which has a temperature
between about 600 and 700 K.  As noted in Ref. \onlinecite{boivineau01}, ``the delocalization process of the $5f$ electrons,''
i.e., the correlation effects, continues to produce anomalies as high as 2000 K in temperature in
many of these properties, well into the high-temperature liquid phase of Pu.  Also, see Ref. \onlinecite{boivineau09}.}
In this regard, it should be noted that although temperature certainly plays an important role
in predicting the correct equilibrium crystal structure,
we believe that it is the resulting volume per atom of
any Pu phase that determines the amount of correlation,
since this is an electronic property.
In particular, we do not expect that
the bandwidth predicted by our zero-temperature $GW$ calculations will be sensitive to any temperature
in the range set by the Pu solid phases.  

The choice of bandwidth narrowing as a measure of correlation strength
is consistent with ideas of correlation going back almost to the beginning
of modern electronic structure theory.  Quasiparticle descriptions
of electronic structure have been standard since Landau developed
Fermi liquid theory and have been derived from
standard many-body approaches (see, for example, the discussion in
Refs.~\onlinecite{abrikosov63,nozieres64,hedin69}).
They have since been extended to strongly correlated electronic
materials (see, for example, the review in Ref.~\onlinecite{hewson93}).
Much of our modern understanding of correlation effects has been
developed using simple model Hamiltonians, especially 
the Hubbard model.\cite{hubbard63}
For metals, most of these approaches for strong correlations
have focused on low-temperatures,\cite{hewson93}
where the electronic structure at the Fermi energy can yield
a rich and diverse set of phenomena at low-energy scales.
In such a case, for example, specific heat or effective
mass enhancements at the Fermi energy have often
been used to characterize the strength of correlations.
As we describe below, pure elemental plutonium forms
correlated states at very high temperatures, and therefore
electronic states are sampled that are far from the Fermi
energy.  Although it is an interesting question how far
away from the Fermi energy correlations effects extend
(see, e.g., Ref.~\onlinecite{byczuk07}),
it is nonetheless important to include correlation
effects for all the quasiparticle states of the $f$ electrons
in Pu.  By including the real part of the self-energy
for all of these states, which are involved in the band
narrowing, our $GW$ approach is thus more relevant for these
high-temperature correlated phases than more
traditional measures of correlation that focus exclusively
on effects at or near the Fermi energy.

To set an appropriate correlation scale, we define our theoretical $C$ by
\begin{equation}
C = 1 - {\textrm w_{rel}},
\label{eqn:c_theor}
\end{equation}
which ranges from $C=0$ (no bandwidth reduction) in the LDA limit to $C=1$ in
the fully localized or atomic limit (the bandwidth becomes zero).

As mentioned above, our test case for correlation is elemental Pu,
an actinide metal, which exhibits large volume changes 
compared to predictions from  band structure theory that are clearly due to correlation 
effects.\cite{wick67,hecker01,hecker04a,hecker04b,albers01}
The large variation in volumes is
controlled by the amount of strong $f$-bonding, which is due to direct $f$-$f$ wave-function overlap.  
The $f$ bonding for many of the different phases is greatly reduced leading to anomalous volume expansions due
to the narrowing of the $f$ bands that results from correlation effects.~\cite{albers01}  
If no correlation were present, the $f$ bonds would
have their full strength and a relatively small volume per atom for all phases 
would be accurately predicted by LDA band-structure methods.  
In the limit of extremely strong correlation the bands would have narrowed so much 
that the $f$ electrons would be fully localized, and they would not contribute to the bonding.  
The volume per atom would then be much larger and close to that of Am, 
which has fully localized $f$ electrons that do not extend outside the atomic core. 

Using the QSGW approximation we have 
calculated~\footnote{We have not included spin-orbit effects, which can be safely ignored for the purposes of this paper. 
The Pu $f$ DOS  splits into a pair of clearly separated $j=5/2$ and $7/2$ peaks.  
To include spin-orbit, we would need to calculate the bandwidth of each peak separately 
and use that corresponding to $j=5/2$.  
By ignoring spin-orbit coupling, we are saved from this additional trouble, which is not expected to 
change the effective $f$ bandwidths.
Recent spin-orbit $GW$ calculations have been calculated 
in Pu (Ref.~\onlinecite{kutepov12}). 
However these have been done in the fully self-consistent
$GW$ method, which usually is a poor approximation in solids due to an incorrect treatment of plasmon effects.
Since the DOS in this paper includes broadening effects due to the imaginary part of the self-energy
in all of the different approximation that were used,
it is also unclear how bandwidth narrowing would separately be affected by spin-orbit effects.}  
the quasiparticle band structures of the fcc, bcc, simple cubic (sc), $\gamma$, and pseudo-$\alpha$  phases of Pu as a
function of volume.  
The pseudo-$\alpha$ is a two-atom per unit cell approximation~\cite{psalpha} to the true $\alpha$ structure of Pu that
preserves the approximate nearest-neighbor distances and other essential features needed for the electronic-structure. 
In this way we avoid performing an extremely large and expensive 16-atom per unit cell calculation for the $\alpha$ structure. 
We are unfortunately unable to present $GW$ results for the $\beta$ structure,
which is even more complex than the $\alpha$ structure, since no pseudostructure for this crystal
structure is available and a QSGW calculation is presently not feasible for so many atoms per unit cell.

To calculate the $f$-electron bandwidths from the $f$-electron projected density of states (DOS), D$_f$(E),
an algorithm is needed to determine the width of the main peak in this DOS.
A simple first guess is to choose a rectangular DOS and to use a least-squares fit to the $GW$
or LDA $f$-DOS to determine the best height and width of the rectangle.  
A drawback of this method is that an artificial broadening of the effective $f$ bandwidth appears,
which is due to a significant $d$-$f$ hybridization at the bottom of the $f$-DOS that creates  an extra peak at low energies.  
This masks the correlation-induced band narrowing.  Since this peak has relatively lower height than the main $f$ peak, 
we may avoid this complication by generating an algorithm that emphasizes the ``high-peak'' part of the $f$-DOS.  
The algorithm we have used is therefore the second moment of the $f$ DOS
\begin{equation}
W = 2 ( \langle E^2 \rangle - \langle E \rangle^2 )^{1/2}.
\label{eqn:band-width}
\end{equation} 
The factor of two is needed because the bandwidth extends above and below the mean energy 
and is not just the average deviation from the mean energy.  
To emphasize the main part of the $f$-DOS peak, the square of the $f$ DOS is used 
as weight function:\footnote{Choosing instead for example $D_{f}(E)$ as the weight function does not serve the 
purpose of emphasizing the central over the ``hybridization wings'' in the $f$ DOS.}
\begin{equation}
\langle f(E)\rangle  \equiv \int dE f(E) D_f^2(E) / \int dE D_f^2(E).
\label{eqn:weight}
\end{equation}

\section{Numerical Results and Discussion}
\label{sec:results}

\begin{figure}
\includegraphics[width=0.95\columnwidth,trim = 0.1in 0.0in 0.4in 0.0in,clip]{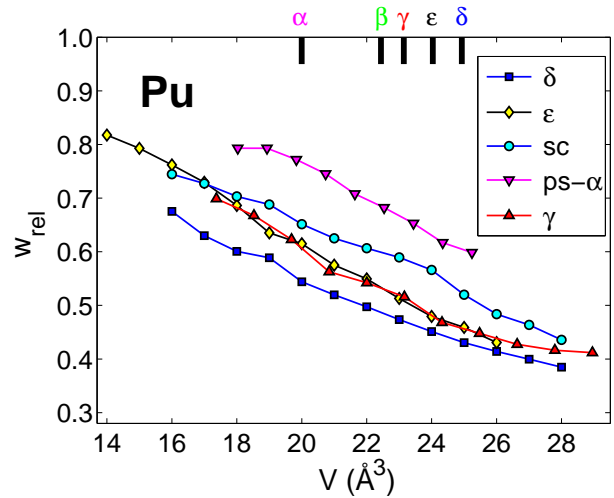}
\caption{\label{fig:pu_bwred_V} (Color online) Plot of w$_{rel}$= $W_f$(GW)/$W_f$(LDA) versus volume, $V$, per atom,
 for the $\gamma$, fcc, bcc, sc, and ps-$\alpha$
 [pseudo-$\alpha$, an approximate $\alpha$-phasen (Ref.~\onlinecite{psalpha})] crystal phases of Pu. 
Note that the sc (simple cubic) is a hypothetical structure for Pu.
 The small, vertical bars at the top of the figure mark the experimentally observed atomic 
volumes (Ref.~54).
}
\end{figure}

In Fig.~\ref{fig:pu_bwred_V} 
we illustrate how  w$_{rel}$ varies with volume for the five different phases considered 
here.~\footnote{For the atomic volumes we have ignored any thermal volume expansion.  
Each phase is represented by a volume corresponding to a fixed temperature within that phase.  
We have used the original data of Zachariasen and 
Ellinger (Refs.~\onlinecite{zachariasen63b,zachariasen63a,zachariasen55,ellinger56}) 
corresponding to the volumes at the temperatures 21, 190, 235, 320, 477, and 490 $^\circ$C, 
for the $\alpha$, $\beta$, $\gamma$, $\delta$, $\delta'$, and $\epsilon$ phases, respectively.}
Large volume variations ranging  between about 14--28 \AA$^3$ per atom are considered, 
with bandwidths that span almost an order of magnitude, 
from about 0.5 eV to 2.5 eV. 
Although the LDA bandwidth decreases with increased volume due to reduction 
in $f$-$f$ overlap of the wavefunctions, 
the QSGW bandwidth decreases 
even faster illustrating increased correlation effects with lattice expansion.  
The bandwidth at a specific volume depends on crystal structure (due to differences in coordination and bond lengths),
as does also the correlation strength.

\begin{figure}
\includegraphics[width=0.95\columnwidth,trim = 1.5in 3.4in 1.8in 3.3in, clip]{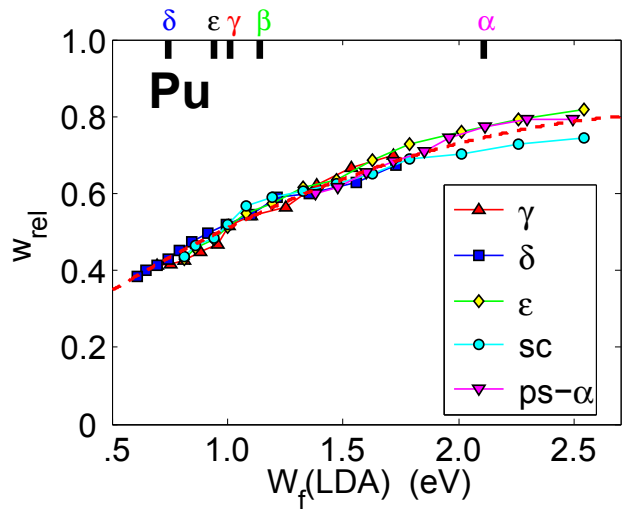}
\caption{\label{fig:pu_bw_reduc} (Color online) Plot of w$_{rel}$= $W_f$(GW)/$W_f$(LDA) versus $W_f$(LDA)
 for the $\gamma$, fcc, bcc, sc, and ps-$\alpha$.
The dashed red line represents the fit of Eq.~(\ref{eqn:w_rel_fit})
The small, vertical bars at the top of the figure mark the values of $W_f$(LDA) calculated 
at the experimental volumes of the five Pu phases (Ref.~54).
}
\end{figure}

Although we expect electronic-structure calculations to strongly depend on the crystal structure 
and lattice constant, we surprisingly found that correlation
effects were approximately independent of these.
Indeed, Fig.~\ref{fig:pu_bw_reduc} shows that all of
our different calculations for our measure of correlation strength,
the reduced bandwidth, collapse to a single ``universal" curve
when plotted as a function of the LDA bandwidth.
In making this plot, it is likely that the effective screened 
Coulomb interaction between the $5f$ electrons is 
approximately constant and that the correlation effects
are being tuned by the effective average kinetic energy of these
electrons as reflected in their LDA bandwidth.
In the range of $W_f$ values considered here the curve is approximately quadratic, i.e., 
\begin{equation}
{\textrm w_{rel}}(x) = 0.15 + 0.43x -0.07x^2,
\label{eqn:w_rel_fit}
\end{equation}
where $x$ = $W_f$(LDA) in eV.  
From Eq. (\ref{eqn:c_theor}) we can use these results to determine a correlation strength $C$. 
It is remarkable that the many-body properties of a strongly correlated system
can be tuned with what is normally considered to be a one-electron property.

\begin{figure}
\includegraphics[width=0.95\columnwidth,trim = 0.0in 2.0in 0.4in 0.2in,clip]{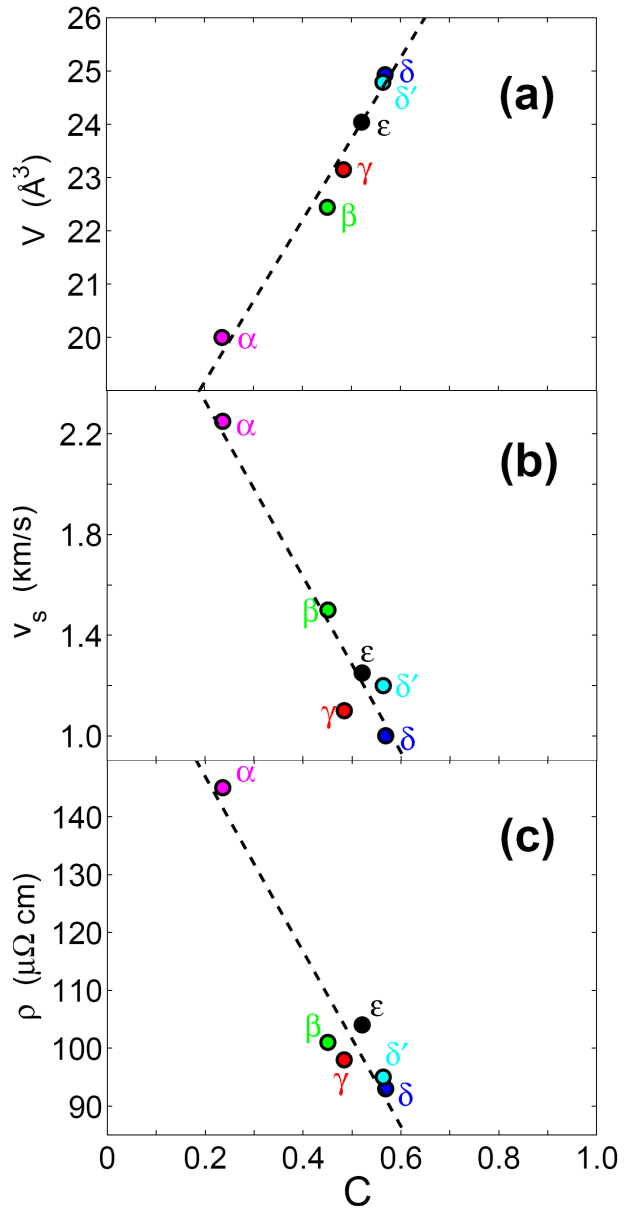}
\caption{\label{fig:trends} (Color online) Trends in Pu properties as a function of correlation strength $C$,
including (a) volume per atom (Ref.~54), (b) sound velocity (Ref.~\onlinecite{boivineau01}), 
and (c) resistivity (Ref.~\onlinecite{boivineau01}).}
\end{figure}

In Fig.~\ref{fig:trends} we 
show~\footnote{For the volumes of the different phases of
Pu, we have followed the same method used to generate Fig.~\ref{fig:pu_bwred_V}.
We have also used the same volumes of the different phases for the sound velocity and resistivity needed
to determine the correlation strength from the $GW$ calculations plotted in Fig.~\ref{fig:pu_bwred_V}.
Note that, since we have not directly calculated the value of ${\textrm w_{rel}}$
for the $\beta$ phase, we instead used the availability of the
bandwidth reduction of Eq. (\ref{eqn:w_rel_fit}) 
together with the calculated LDA bandwidth for the
correct crystal structure of $\beta$ Pu 
to determine ${w_{rel}}(\beta)$ = 0.55.}
that our definition of theoretical correlation strength does indeed
fulfill our expectations and can be used to
bring order into the trends for various experimental properties, including volume, 
sound velocity, and resistivity. 
These properties exhibit an approximately 25\%, 50\%, and 35\% change over the correlation
range (about 0.2 to 0.6) between the $\alpha$ and $\delta$ phases of Pu and, with some scatter
that might partially depend on sample quality, fall on smooth curves when
plotted as a function of our theoretical correlation strength.
It is remarkable that all of these data should collapse to a single curve
for each property that is independent of any explicit consideration of
temperature, crystal structure, or other variable.
However, more generally, we would only expect this to be true
for a property that was predominantly affected by correlation effects.

In terms of theoretical trends, various theories have often attempted to estimate
the amount of correlation in terms of the $Z$-factor,
\begin{equation}
Z_{n{\mathbf k}} =  \left( 1- \langle\Psi_{n{\mathbf k}}|
\frac{\partial\Sigma(\epsilon_{n{\mathbf k}})}{\partial\omega}
 |\Psi_{n{\mathbf k}}\rangle\right)^{-1},
\label{eqn:Z}
\end{equation}
where $\Psi_{n{\mathbf k}}$ are the (LDA) electronic eigenfunctions with energies $\epsilon_{n{\mathbf k}}$, 
and $\Sigma$ denotes the self-energy.
We have found that the volume dependence of the $Z$-factors follows the trend of the $f$-bandwidth reduction 
in Fig.~\ref{fig:pu_bwred_V}, i.e., our measure of correlation strength, 
albeit with variations due to ${\mathbf k}$- and hybridization-dependence. 
However, it should be noted that the relation between $Z$ and bandwidth
reduction is not the same in all materials, especially for weakly correlated broad-band systems,
which seem very different from strongly correlated materials such as Pu.

\begin{figure}
\includegraphics[width=0.98\columnwidth,trim = 1.4in 3.4in 1.6in 3.4in, clip]{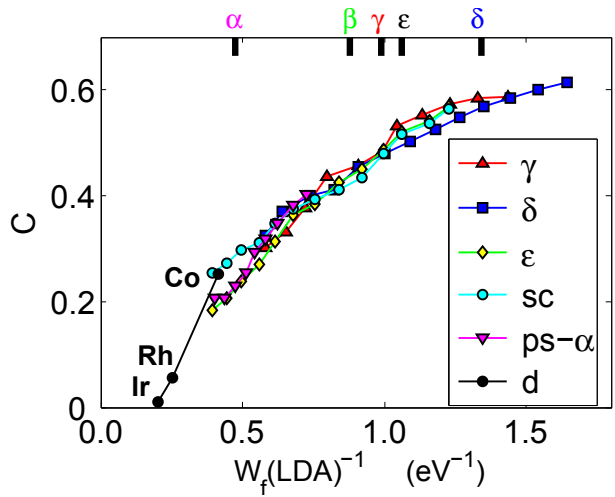}
\caption{\label{fig:one_over_w} (Color online) $C$ from $GW$ theory versus 1/W$_{f}(LDA)$. 
The data for Co, Rh and Ir are for the $3d$, $4d$, and $5d$ bandwidths, respectively. 
The small, vertical bars at the top of the figure mark the values of W$_{f}$(LDA)$^{-1}$ calculated
at the experimental volumes of the five Pu phases (Ref.~54).}
\end{figure}

The simplest Hubbard-like Hamiltonian\cite{hubbard63} to describe
strongly correlated electron systems has a form
\begin{equation}
H = \sum_{ij,\sigma} t_{ij} c^{\dagger}_{i\sigma} c_{j\sigma} + U \sum_{i} n_{i\uparrow} n_{i\downarrow} .
\label{eqn:hubbard}
\end{equation}
with two parameters: the Hubbard parameter $U$ which induces correlation, 
and an effective $t$, which can be related to the {\em uncorrelated} bandwidth $W$.  
When $W$ dominates, the system is in a weakly correlated limit and, when $U$ dominates, the system is in a strongly correlated
regime.  Hence, one can study the solutions as a function of $U/W$ to go from one limit to another.  
In more realistic electronic-structure calculations, the same physics is intuitively expected to carry over.  
The Hubbard $U$ can then be thought of as a screened on-site Coulomb interaction
and the bandwidth as due to the normal band-structure hybridization. 
In our context, this suggests that the correlation strength $C$ should also be a
function of $U/W$.  
To test this, in Fig.~\ref{fig:one_over_w} we plot $C$ versus $1/W_f(\text{LDA})$.
If the effective $U$ were approximately constant, we had hoped to observe
some approximate linear behavior at weak correlations, but any
such behavior is unclear in Fig.~\ref{fig:one_over_w}.
To show what might happen at weaker correlation strengths we
have also included in Fig.~\ref{fig:one_over_w} the equilibrium-volume results 
for Co, Rh, and Ir for the $d$-electron projected DOS.
Interestingly enough, the $d$-electron results seem to follow the same overall trend to large
bandwidths (small correlation).
Among the transition metals included in the plot, Co (3$d$) has the most narrow $d$ band, 
and the correlation value is close to the lowest values for Pu in the figure.

\section{Conclusion}
\label{sec:summary}

In summary, we have introduced the idea of a ``correlation strength" quantity $C$,
which must be taken into account in order to explain the properties of strongly correlated electronic materials.
As an example, we have shown how to use the $GW$ method to define a theoretical $C$ for metallic Pu,
and that various experimental physical properties, including anomalous volume expansion, sound velocity, 
and resistivity, for the different phases of Pu follow well-defined trends when plotted versus our theoretical correlation strength.  
We have also demonstrated a universal scaling relationship for the correlation-reduced bandwidth as a function of the LDA bandwidth.  

\begin{acknowledgments}
This work was carried out under the auspices of the National  Nuclear Security Administration
 of the U.S. Department of Energy  at Los Alamos National Laboratory under Contract No. DE-AC52-06NA25396, 
the Los Alamos LDRD Program,
and the Research Foundation of Aarhus University.
The calculations were carried out at the Centre for Scientific Computing in Aarhus (CSC-AA), financed by 
the Danish Centre for Scientific Computing (DCSC) and the Faculty of Science and Technology, Aarhus University.
\end{acknowledgments}

\bibliographystyle{apsrev4-1}

\begin{thebibliography}{62}%
\makeatletter
\providecommand \@ifxundefined [1]{%
 \@ifx{#1\undefined}
}%
\providecommand \@ifnum [1]{%
 \ifnum #1\expandafter \@firstoftwo
 \else \expandafter \@secondoftwo
 \fi
}%
\providecommand \@ifx [1]{%
 \ifx #1\expandafter \@firstoftwo
 \else \expandafter \@secondoftwo
 \fi
}%
\providecommand \natexlab [1]{#1}%
\providecommand \enquote  [1]{``#1''}%
\providecommand \bibnamefont  [1]{#1}%
\providecommand \bibfnamefont [1]{#1}%
\providecommand \citenamefont [1]{#1}%
\providecommand \href@noop [0]{\@secondoftwo}%
\providecommand \href [0]{\begingroup \@sanitize@url \@href}%
\providecommand \@href[1]{\@@startlink{#1}\@@href}%
\providecommand \@@href[1]{\endgroup#1\@@endlink}%
\providecommand \@sanitize@url [0]{\catcode `\\12\catcode `\$12\catcode
  `\&12\catcode `\#12\catcode `\^12\catcode `\_12\catcode `\%12\relax}%
\providecommand \@@startlink[1]{}%
\providecommand \@@endlink[0]{}%
\providecommand \url  [0]{\begingroup\@sanitize@url \@url }%
\providecommand \@url [1]{\endgroup\@href {#1}{\urlprefix }}%
\providecommand \urlprefix  [0]{URL }%
\providecommand \Eprint [0]{\href }%
\providecommand \doibase [0]{http://dx.doi.org/}%
\providecommand \selectlanguage [0]{\@gobble}%
\providecommand \bibinfo  [0]{\@secondoftwo}%
\providecommand \bibfield  [0]{\@secondoftwo}%
\providecommand \translation [1]{[#1]}%
\providecommand \BibitemOpen [0]{}%
\providecommand \bibitemStop [0]{}%
\providecommand \bibitemNoStop [0]{.\EOS\space}%
\providecommand \EOS [0]{\spacefactor3000\relax}%
\providecommand \BibitemShut  [1]{\csname bibitem#1\endcsname}%
\let\auto@bib@innerbib\@empty
\bibitem [{\citenamefont {Hedin}(1965)}]{hedin65}%
  \BibitemOpen
  \bibfield  {author} {\bibinfo {author} {\bibfnamefont {L.}~\bibnamefont
  {Hedin}},\ }\href@noop {} {\bibfield  {journal} {\bibinfo  {journal} {Phys.
  Rev.}\ }\textbf {\bibinfo {volume} {139}},\ \bibinfo {pages} {A796} (\bibinfo
  {year} {1965})}\BibitemShut {NoStop}%
\bibitem [{\citenamefont {Hedin}\ and\ \citenamefont
  {Lundqvist}(1969)}]{hedin69}%
  \BibitemOpen
  \bibfield  {author} {\bibinfo {author} {\bibfnamefont {L.}~\bibnamefont
  {Hedin}}\ and\ \bibinfo {author} {\bibfnamefont {S.}~\bibnamefont
  {Lundqvist}},\ }in\ \href@noop {} {\emph {\bibinfo {booktitle} {Solid State
  Physics}}},\ Vol.~\bibinfo {volume} {23},\ \bibinfo {editor} {edited by\
  \bibinfo {editor} {\bibfnamefont {F.}~\bibnamefont {Seitz}}, \bibinfo
  {editor} {\bibfnamefont {D.}~\bibnamefont {Turnbull}}, \ and\ \bibinfo
  {editor} {\bibfnamefont {H.}~\bibnamefont {Ehrenreich}}}\ (\bibinfo
  {publisher} {Academic Press},\ \bibinfo {address} {New York},\ \bibinfo
  {year} {1969})\ pp.\ \bibinfo {pages} {1--181}\BibitemShut {NoStop}%
\bibitem [{\citenamefont {Hedin}(1999)}]{hedin99}%
  \BibitemOpen
  \bibfield  {author} {\bibinfo {author} {\bibfnamefont {L.}~\bibnamefont
  {Hedin}},\ }\href@noop {} {\bibfield  {journal} {\bibinfo  {journal} {J.
  Phys.: Condens. Matter}\ }\textbf {\bibinfo {volume} {11}},\ \bibinfo {pages}
  {R489} (\bibinfo {year} {1999})}\BibitemShut {NoStop}%
\bibitem [{\citenamefont {Aryasetiawan}\ and\ \citenamefont
  {Gunnarsson}(1998)}]{aryasetiwan98}%
  \BibitemOpen
  \bibfield  {author} {\bibinfo {author} {\bibfnamefont {F.}~\bibnamefont
  {Aryasetiawan}}\ and\ \bibinfo {author} {\bibfnamefont {O.}~\bibnamefont
  {Gunnarsson}},\ }\href@noop {} {\bibfield  {journal} {\bibinfo  {journal}
  {Rep. Prog. Phys.}\ }\textbf {\bibinfo {volume} {61}},\ \bibinfo {pages}
  {247} (\bibinfo {year} {1998})}\BibitemShut {NoStop}%
\bibitem [{\citenamefont {Chantis}\ \emph {et~al.}(2009)\citenamefont
  {Chantis}, \citenamefont {Albers}, \citenamefont {Svane},\ and\ \citenamefont
  {Christensen}}]{chantis09}%
  \BibitemOpen
  \bibfield  {author} {\bibinfo {author} {\bibfnamefont {A.~N.}\ \bibnamefont
  {Chantis}}, \bibinfo {author} {\bibfnamefont {R.~C.}\ \bibnamefont {Albers}},
  \bibinfo {author} {\bibfnamefont {A.}~\bibnamefont {Svane}}, \ and\ \bibinfo
  {author} {\bibfnamefont {N.~E.}\ \bibnamefont {Christensen}},\ }\href@noop {}
  {\bibfield  {journal} {\bibinfo  {journal} {Phil. Mag.}\ }\textbf {\bibinfo
  {volume} {89}},\ \bibinfo {pages} {1801} (\bibinfo {year}
  {2009})}\BibitemShut {NoStop}%
\bibitem [{\citenamefont {Kutepov}\ \emph {et~al.}(2012)\citenamefont
  {Kutepov}, \citenamefont {Haule}, \citenamefont {Savrasov},\ and\
  \citenamefont {Kotliar}}]{kutepov12}%
  \BibitemOpen
  \bibfield  {author} {\bibinfo {author} {\bibfnamefont {A.}~\bibnamefont
  {Kutepov}}, \bibinfo {author} {\bibfnamefont {K.}~\bibnamefont {Haule}},
  \bibinfo {author} {\bibfnamefont {S.~Y.}\ \bibnamefont {Savrasov}}, \ and\
  \bibinfo {author} {\bibfnamefont {G.}~\bibnamefont {Kotliar}},\ }\href@noop
  {} {\bibfield  {journal} {\bibinfo  {journal} {Phys.\ Rev.\ B}\ }\textbf
  {\bibinfo {volume} {85}},\ \bibinfo {pages} {155129} (\bibinfo {year}
  {2012})}\BibitemShut {NoStop}%
\bibitem [{\citenamefont {Hill}(1970)}]{hill70}%
  \BibitemOpen
  \bibfield  {author} {\bibinfo {author} {\bibfnamefont {H.~H.}\ \bibnamefont
  {Hill}},\ }in\ \href@noop {} {\emph {\bibinfo {booktitle} {Plutonium 1970 and
  Other Actinides}}},\ \bibinfo {editor} {edited by\ \bibinfo {editor}
  {\bibfnamefont {W.~N.}\ \bibnamefont {Miner}}}\ (\bibinfo  {publisher}
  {American Institute of Mechanical Engineers},\ \bibinfo {address} {New
  York},\ \bibinfo {year} {1970})\ p.~\bibinfo {pages} {2}\BibitemShut
  {NoStop}%
\bibitem [{\citenamefont {Boring}\ and\ \citenamefont
  {Smith}(2000)}]{boring2000}%
  \BibitemOpen
  \bibfield  {author} {\bibinfo {author} {\bibfnamefont {A.~M.}\ \bibnamefont
  {Boring}}\ and\ \bibinfo {author} {\bibfnamefont {J.~L.}\ \bibnamefont
  {Smith}},\ }in\ \href@noop {} {\emph {\bibinfo {booktitle} {Challenges in
  Plutonium Science}}},\ \bibinfo {series and number} {\bibinfo {series} {Los
  Alamos Science}\ No.~\bibinfo {number} {26}},\ \bibinfo {editor} {edited by\
  \bibinfo {editor} {\bibfnamefont {N.~G.}\ \bibnamefont {Cooper}}}\ (\bibinfo
  {publisher} {Los Alamos National Laboratory},\ \bibinfo {address} {Los
  Alamos},\ \bibinfo {year} {2000})\ pp.\ \bibinfo {pages}
  {90--127}\BibitemShut {NoStop}%
\bibitem [{\citenamefont {Smith}\ and\ \citenamefont {Kmetko}(1983)}]{smith83}%
  \BibitemOpen
  \bibfield  {author} {\bibinfo {author} {\bibfnamefont {J.~L.}\ \bibnamefont
  {Smith}}\ and\ \bibinfo {author} {\bibfnamefont {E.~A.}\ \bibnamefont
  {Kmetko}},\ }\href@noop {} {\bibfield  {journal} {\bibinfo  {journal} {J.
  Less Common Met.}\ }\textbf {\bibinfo {volume} {90}},\ \bibinfo {pages} {83}
  (\bibinfo {year} {1983})}\BibitemShut {NoStop}%
\bibitem [{\citenamefont {Pettifor}(2000)}]{pettifor00}%
  \BibitemOpen
  \bibfield  {author} {\bibinfo {author} {\bibfnamefont {D.~G.}\ \bibnamefont
  {Pettifor}},\ }\enquote {\bibinfo {title} {Structure mapping},}\ in\
  \href@noop {} {\emph {\bibinfo {booktitle} {Structures of Intermetallic
  Compounds}}},\ \bibinfo {editor} {edited by\ \bibinfo {editor} {\bibfnamefont
  {J.~H.}\ \bibnamefont {Westbrook}}\ and\ \bibinfo {editor} {\bibfnamefont
  {R.~L.}\ \bibnamefont {Fleischer}}}\ (\bibinfo  {publisher} {Wiley},\
  \bibinfo {address} {New York},\ \bibinfo {year} {2000})\ Chap.~\bibinfo
  {chapter} {8}, pp.\ \bibinfo {pages} {195--214}\BibitemShut {NoStop}%
\bibitem [{\citenamefont {Villars}(1983)}]{villars83}%
  \BibitemOpen
  \bibfield  {author} {\bibinfo {author} {\bibfnamefont {P.}~\bibnamefont
  {Villars}},\ }\href@noop {} {\bibfield  {journal} {\bibinfo  {journal} {J.
  Less Common Met.}\ }\textbf {\bibinfo {volume} {92}},\ \bibinfo {pages} {215}
  (\bibinfo {year} {1983})}\BibitemShut {NoStop}%
\bibitem [{\citenamefont {Pettifor}(1986)}]{pettifor86}%
  \BibitemOpen
  \bibfield  {author} {\bibinfo {author} {\bibfnamefont {D.~G.}\ \bibnamefont
  {Pettifor}},\ }\href@noop {} {\bibfield  {journal} {\bibinfo  {journal} {J.
  Phys. C}\ }\textbf {\bibinfo {volume} {19}},\ \bibinfo {pages} {285}
  (\bibinfo {year} {1986})}\BibitemShut {NoStop}%
\bibitem [{\citenamefont {Christensen}\ \emph {et~al.}(1987)\citenamefont
  {Christensen}, \citenamefont {Satpathy},\ and\ \citenamefont
  {Pawlowska}}]{christensen87}%
  \BibitemOpen
  \bibfield  {author} {\bibinfo {author} {\bibfnamefont {N.~E.}\ \bibnamefont
  {Christensen}}, \bibinfo {author} {\bibfnamefont {S.}~\bibnamefont
  {Satpathy}}, \ and\ \bibinfo {author} {\bibfnamefont {Z.}~\bibnamefont
  {Pawlowska}},\ }\href@noop {} {\bibfield  {journal} {\bibinfo  {journal}
  {Phys. Rev. B}\ }\textbf {\bibinfo {volume} {36}},\ \bibinfo {pages} {1032}
  (\bibinfo {year} {1987})}\BibitemShut {NoStop}%
\bibitem [{\citenamefont {Pettifor}(1988)}]{pettifor88}%
  \BibitemOpen
  \bibfield  {author} {\bibinfo {author} {\bibfnamefont {D.~G.}\ \bibnamefont
  {Pettifor}},\ }\href@noop {} {\bibfield  {journal} {\bibinfo  {journal}
  {Materials Science and Technology}\ }\textbf {\bibinfo {volume} {4}},\
  \bibinfo {pages} {675} (\bibinfo {year} {1988})}\BibitemShut {NoStop}%
\bibitem [{\citenamefont {Fischer}\ \emph {et~al.}(2006)\citenamefont
  {Fischer}, \citenamefont {Tibbetts}, \citenamefont {Morgan},\ and\
  \citenamefont {Ceder}}]{fischer06}%
  \BibitemOpen
  \bibfield  {author} {\bibinfo {author} {\bibfnamefont {C.~C.}\ \bibnamefont
  {Fischer}}, \bibinfo {author} {\bibfnamefont {K.~J.}\ \bibnamefont
  {Tibbetts}}, \bibinfo {author} {\bibfnamefont {D.}~\bibnamefont {Morgan}}, \
  and\ \bibinfo {author} {\bibfnamefont {G.}~\bibnamefont {Ceder}},\
  }\href@noop {} {\bibfield  {journal} {\bibinfo  {journal} {Nature Materials}\
  }\textbf {\bibinfo {volume} {5}},\ \bibinfo {pages} {641} (\bibinfo {year}
  {2006})}\BibitemShut {NoStop}%
\bibitem [{\citenamefont {Fischer}\ \emph {et~al.}(2007)\citenamefont
  {Fischer}, \citenamefont {Kugler}, \citenamefont {Maggio-Aprile},\ and\
  \citenamefont {Berthod}}]{fischer07}%
  \BibitemOpen
  \bibfield  {author} {\bibinfo {author} {\bibfnamefont {O.}~\bibnamefont
  {Fischer}}, \bibinfo {author} {\bibfnamefont {M.}~\bibnamefont {Kugler}},
  \bibinfo {author} {\bibfnamefont {I.}~\bibnamefont {Maggio-Aprile}}, \ and\
  \bibinfo {author} {\bibfnamefont {C.}~\bibnamefont {Berthod}},\ }\href@noop
  {} {\bibfield  {journal} {\bibinfo  {journal} {Rev. Mod. Phys.}\ }\textbf
  {\bibinfo {volume} {79}},\ \bibinfo {pages} {353} (\bibinfo {year}
  {2007})}\BibitemShut {NoStop}%
\bibitem [{\citenamefont {Takahashi}\ \emph {et~al.}(2008)\citenamefont
  {Takahashi}, \citenamefont {Igawa}, \citenamefont {Arii}, \citenamefont
  {Kamihara}, \citenamefont {Hirano},\ and\ \citenamefont
  {Hosono}}]{takahashi08}%
  \BibitemOpen
  \bibfield  {author} {\bibinfo {author} {\bibfnamefont {H.}~\bibnamefont
  {Takahashi}}, \bibinfo {author} {\bibfnamefont {K.}~\bibnamefont {Igawa}},
  \bibinfo {author} {\bibfnamefont {K.}~\bibnamefont {Arii}}, \bibinfo {author}
  {\bibfnamefont {Y.}~\bibnamefont {Kamihara}}, \bibinfo {author}
  {\bibfnamefont {M.}~\bibnamefont {Hirano}}, \ and\ \bibinfo {author}
  {\bibfnamefont {H.}~\bibnamefont {Hosono}},\ }\href@noop {} {\bibfield
  {journal} {\bibinfo  {journal} {Nature (London)}\ }\textbf {\bibinfo {volume}
  {453}},\ \bibinfo {pages} {376} (\bibinfo {year} {2008})}\BibitemShut
  {NoStop}%
\bibitem [{\citenamefont {Bauer}\ \emph {et~al.}(2004)\citenamefont {Bauer},
  \citenamefont {Thompson}, \citenamefont {Sarrao}, \citenamefont {Morales},
  \citenamefont {F.Wastin}, \citenamefont {Rebizant}, \citenamefont {Griveau},
  \citenamefont {Javorsky}, \citenamefont {Boulet}, \citenamefont {Colineau},
  \citenamefont {Lander}, ,\ and\ \citenamefont {Stewart}}]{bauer04}%
  \BibitemOpen
  \bibfield  {author} {\bibinfo {author} {\bibfnamefont {E.~D.}\ \bibnamefont
  {Bauer}}, \bibinfo {author} {\bibfnamefont {J.~D.}\ \bibnamefont {Thompson}},
  \bibinfo {author} {\bibfnamefont {J.~L.}\ \bibnamefont {Sarrao}}, \bibinfo
  {author} {\bibfnamefont {L.~A.}\ \bibnamefont {Morales}}, \bibinfo {author}
  {\bibnamefont {F.Wastin}}, \bibinfo {author} {\bibfnamefont {J.}~\bibnamefont
  {Rebizant}}, \bibinfo {author} {\bibfnamefont {J.~C.}\ \bibnamefont
  {Griveau}}, \bibinfo {author} {\bibfnamefont {P.}~\bibnamefont {Javorsky}},
  \bibinfo {author} {\bibfnamefont {P.}~\bibnamefont {Boulet}}, \bibinfo
  {author} {\bibfnamefont {E.}~\bibnamefont {Colineau}}, \bibinfo {author}
  {\bibfnamefont {G.~H.}\ \bibnamefont {Lander}}, , \ and\ \bibinfo {author}
  {\bibfnamefont {G.~R.}\ \bibnamefont {Stewart}},\ }\href@noop {} {\bibfield
  {journal} {\bibinfo  {journal} {Phys.\ Rev.\ Lett.}\ }\textbf {\bibinfo
  {volume} {93}},\ \bibinfo {pages} {147005} (\bibinfo {year}
  {2004})}\BibitemShut {NoStop}%
\bibitem [{\citenamefont {Sarrao}\ and\ \citenamefont
  {Thompson}(2007)}]{sarrao07}%
  \BibitemOpen
  \bibfield  {author} {\bibinfo {author} {\bibfnamefont {J.~L.}\ \bibnamefont
  {Sarrao}}\ and\ \bibinfo {author} {\bibfnamefont {J.~D.}\ \bibnamefont
  {Thompson}},\ }\href@noop {} {\bibfield  {journal} {\bibinfo  {journal} {J.
  Phys. Soc. Jpn.}\ }\textbf {\bibinfo {volume} {76}},\ \bibinfo {pages}
  {051013} (\bibinfo {year} {2007})}\BibitemShut {NoStop}%
\bibitem [{\citenamefont {Pfleiderer}(2009)}]{pfleiderer09}%
  \BibitemOpen
  \bibfield  {author} {\bibinfo {author} {\bibfnamefont {C.}~\bibnamefont
  {Pfleiderer}},\ }\href@noop {} {\bibfield  {journal} {\bibinfo  {journal}
  {Rev. Mod. Phys.}\ }\textbf {\bibinfo {volume} {81}},\ \bibinfo {pages}
  {1551} (\bibinfo {year} {2009})}\BibitemShut {NoStop}%
\bibitem [{\citenamefont {Duthie}\ and\ \citenamefont
  {Pettifor}(1977)}]{duthie77}%
  \BibitemOpen
  \bibfield  {author} {\bibinfo {author} {\bibfnamefont {J.~C.}\ \bibnamefont
  {Duthie}}\ and\ \bibinfo {author} {\bibfnamefont {D.~G.}\ \bibnamefont
  {Pettifor}},\ }\href@noop {} {\bibfield  {journal} {\bibinfo  {journal}
  {Phys.\ Rev.\ Lett.}\ }\textbf {\bibinfo {volume} {38}},\ \bibinfo {pages}
  {564} (\bibinfo {year} {1977})}\BibitemShut {NoStop}%
\bibitem [{\citenamefont {Skriver}(1985)}]{skriver85}%
  \BibitemOpen
  \bibfield  {author} {\bibinfo {author} {\bibfnamefont {H.~L.}\ \bibnamefont
  {Skriver}},\ }\href@noop {} {\bibfield  {journal} {\bibinfo  {journal} {Phys.
  Rev. B}\ }\textbf {\bibinfo {volume} {31}},\ \bibinfo {pages} {1909}
  (\bibinfo {year} {1985})}\BibitemShut {NoStop}%
\bibitem [{\citenamefont {Albers}\ \emph {et~al.}(2009)\citenamefont {Albers},
  \citenamefont {Christensen},\ and\ \citenamefont {Svane}}]{albers09}%
  \BibitemOpen
  \bibfield  {author} {\bibinfo {author} {\bibfnamefont {R.~C.}\ \bibnamefont
  {Albers}}, \bibinfo {author} {\bibfnamefont {N.~E.}\ \bibnamefont
  {Christensen}}, \ and\ \bibinfo {author} {\bibfnamefont {A.}~\bibnamefont
  {Svane}},\ }\href@noop {} {\bibfield  {journal} {\bibinfo  {journal} {J.
  Phys.: Condens. Matter}\ }\textbf {\bibinfo {volume} {21}},\ \bibinfo {pages}
  {343201} (\bibinfo {year} {2009})}\BibitemShut {NoStop}%
\bibitem [{\citenamefont {Georges}\ \emph {et~al.}(1996)\citenamefont
  {Georges}, \citenamefont {Kotliar}, \citenamefont {Krauth},\ and\
  \citenamefont {Rozenberg}}]{georges96}%
  \BibitemOpen
  \bibfield  {author} {\bibinfo {author} {\bibfnamefont {A.}~\bibnamefont
  {Georges}}, \bibinfo {author} {\bibfnamefont {G.}~\bibnamefont {Kotliar}},
  \bibinfo {author} {\bibfnamefont {W.}~\bibnamefont {Krauth}}, \ and\ \bibinfo
  {author} {\bibfnamefont {M.~J.}\ \bibnamefont {Rozenberg}},\ }\href@noop {}
  {\bibfield  {journal} {\bibinfo  {journal} {Rev. Mod. Phys.}\ }\textbf
  {\bibinfo {volume} {68}},\ \bibinfo {pages} {13} (\bibinfo {year}
  {1996})}\BibitemShut {NoStop}%
\bibitem [{\citenamefont {Kotliar}\ \emph {et~al.}(2006)\citenamefont
  {Kotliar}, \citenamefont {Savrasov}, \citenamefont {Haule}, \citenamefont
  {Oudovenko}, \citenamefont {Parcollet},\ and\ \citenamefont
  {Marianetti}}]{kotliar06}%
  \BibitemOpen
  \bibfield  {author} {\bibinfo {author} {\bibfnamefont {G.}~\bibnamefont
  {Kotliar}}, \bibinfo {author} {\bibfnamefont {S.~Y.}\ \bibnamefont
  {Savrasov}}, \bibinfo {author} {\bibfnamefont {K.}~\bibnamefont {Haule}},
  \bibinfo {author} {\bibfnamefont {V.~S.}\ \bibnamefont {Oudovenko}}, \bibinfo
  {author} {\bibfnamefont {O.}~\bibnamefont {Parcollet}}, \ and\ \bibinfo
  {author} {\bibfnamefont {C.~A.}\ \bibnamefont {Marianetti}},\ }\href@noop {}
  {\bibfield  {journal} {\bibinfo  {journal} {Rev. Mod. Phys.}\ }\textbf
  {\bibinfo {volume} {78}},\ \bibinfo {pages} {865} (\bibinfo {year}
  {2006})}\BibitemShut {NoStop}%
\bibitem [{\citenamefont {Held}\ \emph {et~al.}(2008)\citenamefont {Held},
  \citenamefont {Andersen}, \citenamefont {Feldbacher}, \citenamefont
  {Yamasaki},\ and\ \citenamefont {Yang}}]{held08}%
  \BibitemOpen
  \bibfield  {author} {\bibinfo {author} {\bibfnamefont {K.}~\bibnamefont
  {Held}}, \bibinfo {author} {\bibfnamefont {O.~K.}\ \bibnamefont {Andersen}},
  \bibinfo {author} {\bibfnamefont {M.}~\bibnamefont {Feldbacher}}, \bibinfo
  {author} {\bibfnamefont {A.}~\bibnamefont {Yamasaki}}, \ and\ \bibinfo
  {author} {\bibfnamefont {Y.-F.}\ \bibnamefont {Yang}},\ }\href@noop {}
  {\bibfield  {journal} {\bibinfo  {journal} {J. Phys.: Condens. Matter}\
  }\textbf {\bibinfo {volume} {20}},\ \bibinfo {pages} {064202} (\bibinfo
  {year} {2008})}\BibitemShut {NoStop}%
\bibitem [{\citenamefont {Kune\v{s}}\ \emph {et~al.}(2010)\citenamefont
  {Kune\v{s}}, \citenamefont {Leonov}, \citenamefont {Kollar}, \citenamefont
  {Byczuk}, \citenamefont {Anisimov},\ and\ \citenamefont
  {Vollhardt}}]{kunes10}%
  \BibitemOpen
  \bibfield  {author} {\bibinfo {author} {\bibfnamefont {J.}~\bibnamefont
  {Kune\v{s}}}, \bibinfo {author} {\bibfnamefont {I.}~\bibnamefont {Leonov}},
  \bibinfo {author} {\bibfnamefont {M.}~\bibnamefont {Kollar}}, \bibinfo
  {author} {\bibfnamefont {K.}~\bibnamefont {Byczuk}}, \bibinfo {author}
  {\bibfnamefont {V.}~\bibnamefont {Anisimov}}, \ and\ \bibinfo {author}
  {\bibfnamefont {D.}~\bibnamefont {Vollhardt}},\ }\href@noop {} {\bibfield
  {journal} {\bibinfo  {journal} {Eur. Phys. J. Special Topics}\ }\textbf
  {\bibinfo {volume} {180}},\ \bibinfo {pages} {5} (\bibinfo {year}
  {2010})}\BibitemShut {NoStop}%
\bibitem [{\citenamefont {Imada}\ and\ \citenamefont {Miyake}(2010)}]{imada10}%
  \BibitemOpen
  \bibfield  {author} {\bibinfo {author} {\bibfnamefont {M.}~\bibnamefont
  {Imada}}\ and\ \bibinfo {author} {\bibfnamefont {T.}~\bibnamefont {Miyake}},\
  }\href@noop {} {\bibfield  {journal} {\bibinfo  {journal} {J. Phys. Soc.
  Jpn.}\ }\textbf {\bibinfo {volume} {79}},\ \bibinfo {pages} {112001}
  (\bibinfo {year} {2010})}\BibitemShut {NoStop}%
\bibitem [{\citenamefont {Sun}\ and\ \citenamefont {Kotliar}(2002)}]{sun02}%
  \BibitemOpen
  \bibfield  {author} {\bibinfo {author} {\bibfnamefont {P.}~\bibnamefont
  {Sun}}\ and\ \bibinfo {author} {\bibfnamefont {G.}~\bibnamefont {Kotliar}},\
  }\href@noop {} {\bibfield  {journal} {\bibinfo  {journal} {Phys. Rev. B}\
  }\textbf {\bibinfo {volume} {66}},\ \bibinfo {pages} {085120} (\bibinfo
  {year} {2002})}\BibitemShut {NoStop}%
\bibitem [{\citenamefont {Biermann}\ \emph {et~al.}(2003)\citenamefont
  {Biermann}, \citenamefont {Aryasetiawan},\ and\ \citenamefont
  {Georges}}]{biermann03}%
  \BibitemOpen
  \bibfield  {author} {\bibinfo {author} {\bibfnamefont {S.}~\bibnamefont
  {Biermann}}, \bibinfo {author} {\bibfnamefont {F.}~\bibnamefont
  {Aryasetiawan}}, \ and\ \bibinfo {author} {\bibfnamefont {A.}~\bibnamefont
  {Georges}},\ }\href@noop {} {\bibfield  {journal} {\bibinfo  {journal}
  {Phys.\ Rev.\ Lett.}\ }\textbf {\bibinfo {volume} {90}},\ \bibinfo {pages}
  {086402} (\bibinfo {year} {2003})}\BibitemShut {NoStop}%
\bibitem [{\citenamefont {Tomczak}\ \emph {et~al.}(2012)\citenamefont
  {Tomczak}, \citenamefont {Casula}, \citenamefont {Miyake}, \citenamefont
  {Aryasetiawan},\ and\ \citenamefont {Biermann}}]{tomczak12}%
  \BibitemOpen
  \bibfield  {author} {\bibinfo {author} {\bibfnamefont {J.~M.}\ \bibnamefont
  {Tomczak}}, \bibinfo {author} {\bibfnamefont {M.}~\bibnamefont {Casula}},
  \bibinfo {author} {\bibfnamefont {T.}~\bibnamefont {Miyake}}, \bibinfo
  {author} {\bibfnamefont {F.}~\bibnamefont {Aryasetiawan}}, \ and\ \bibinfo
  {author} {\bibfnamefont {S.}~\bibnamefont {Biermann}},\ }\href@noop {}
  {\bibfield  {journal} {\bibinfo  {journal} {EPL}\ }\textbf {\bibinfo {volume}
  {100}},\ \bibinfo {pages} {67001} (\bibinfo {year} {2012})}\BibitemShut
  {NoStop}%
\bibitem [{\citenamefont {van Schilfgaarde}\ \emph
  {et~al.}(2006{\natexlab{a}})\citenamefont {van Schilfgaarde}, \citenamefont
  {Kotani},\ and\ \citenamefont {Faleev}}]{vanschilfgaarde06a}%
  \BibitemOpen
  \bibfield  {author} {\bibinfo {author} {\bibfnamefont {M.}~\bibnamefont {van
  Schilfgaarde}}, \bibinfo {author} {\bibfnamefont {T.}~\bibnamefont {Kotani}},
  \ and\ \bibinfo {author} {\bibfnamefont {S.}~\bibnamefont {Faleev}},\
  }\href@noop {} {\bibfield  {journal} {\bibinfo  {journal} {Phys.\ Rev.\
  Lett.}\ }\textbf {\bibinfo {volume} {96}},\ \bibinfo {pages} {226402}
  (\bibinfo {year} {2006}{\natexlab{a}})}\BibitemShut {NoStop}%
\bibitem [{\citenamefont {van Schilfgaarde}\ \emph
  {et~al.}(2006{\natexlab{b}})\citenamefont {van Schilfgaarde}, \citenamefont
  {Kotani},\ and\ \citenamefont {Faleev}}]{vanschilfgaarde06b}%
  \BibitemOpen
  \bibfield  {author} {\bibinfo {author} {\bibfnamefont {M.}~\bibnamefont {van
  Schilfgaarde}}, \bibinfo {author} {\bibfnamefont {T.}~\bibnamefont {Kotani}},
  \ and\ \bibinfo {author} {\bibfnamefont {S.~V.}\ \bibnamefont {Faleev}},\
  }\href@noop {} {\bibfield  {journal} {\bibinfo  {journal} {Phys.\ Rev.\ B}\
  }\textbf {\bibinfo {volume} {74}},\ \bibinfo {pages} {245125} (\bibinfo
  {year} {2006}{\natexlab{b}})}\BibitemShut {NoStop}%
\bibitem [{\citenamefont {Kotani}\ \emph {et~al.}(2007)\citenamefont {Kotani},
  \citenamefont {van Schilfgaarde},\ and\ \citenamefont {Faleev}}]{kotani07}%
  \BibitemOpen
  \bibfield  {author} {\bibinfo {author} {\bibfnamefont {T.}~\bibnamefont
  {Kotani}}, \bibinfo {author} {\bibfnamefont {M.}~\bibnamefont {van
  Schilfgaarde}}, \ and\ \bibinfo {author} {\bibfnamefont {S.~V.}\ \bibnamefont
  {Faleev}},\ }\href@noop {} {\bibfield  {journal} {\bibinfo  {journal} {Phys.\
  Rev.\ B}\ }\textbf {\bibinfo {volume} {76}},\ \bibinfo {pages} {165106}
  (\bibinfo {year} {2007})}\BibitemShut {NoStop}%
\bibitem [{\citenamefont {Svane}\ \emph
  {et~al.}(2010{\natexlab{a}})\citenamefont {Svane}, \citenamefont
  {Christensen}, \citenamefont {Cardona}, \citenamefont {Chantis},
  \citenamefont {van Schilfgaarde},\ and\ \citenamefont {Kotani}}]{svane10}%
  \BibitemOpen
  \bibfield  {author} {\bibinfo {author} {\bibfnamefont {A.}~\bibnamefont
  {Svane}}, \bibinfo {author} {\bibfnamefont {N.~E.}\ \bibnamefont
  {Christensen}}, \bibinfo {author} {\bibfnamefont {M.}~\bibnamefont
  {Cardona}}, \bibinfo {author} {\bibfnamefont {A.~N.}\ \bibnamefont
  {Chantis}}, \bibinfo {author} {\bibfnamefont {M.}~\bibnamefont {van
  Schilfgaarde}}, \ and\ \bibinfo {author} {\bibfnamefont {T.}~\bibnamefont
  {Kotani}},\ }\href@noop {} {\bibfield  {journal} {\bibinfo  {journal} {Phys.
  Rev. B}\ }\textbf {\bibinfo {volume} {81}},\ \bibinfo {pages} {245120}
  (\bibinfo {year} {2010}{\natexlab{a}})}\BibitemShut {NoStop}%
\bibitem [{\citenamefont {Svane}\ \emph
  {et~al.}(2010{\natexlab{b}})\citenamefont {Svane}, \citenamefont
  {Christensen}, \citenamefont {Gorczyca}, \citenamefont {van Schilfgaarde},
  \citenamefont {Chantis},\ and\ \citenamefont {Kotani}}]{svane10b}%
  \BibitemOpen
  \bibfield  {author} {\bibinfo {author} {\bibfnamefont {A.}~\bibnamefont
  {Svane}}, \bibinfo {author} {\bibfnamefont {N.~E.}\ \bibnamefont
  {Christensen}}, \bibinfo {author} {\bibfnamefont {I.}~\bibnamefont
  {Gorczyca}}, \bibinfo {author} {\bibfnamefont {M.}~\bibnamefont {van
  Schilfgaarde}}, \bibinfo {author} {\bibfnamefont {A.~N.}\ \bibnamefont
  {Chantis}}, \ and\ \bibinfo {author} {\bibfnamefont {T.}~\bibnamefont
  {Kotani}},\ }\href@noop {} {\bibfield  {journal} {\bibinfo  {journal} {Phys.
  Rev. B}\ }\textbf {\bibinfo {volume} {82}},\ \bibinfo {pages} {115102}
  (\bibinfo {year} {2010}{\natexlab{b}})}\BibitemShut {NoStop}%
\bibitem [{\citenamefont {Svane}\ \emph {et~al.}(2011)\citenamefont {Svane},
  \citenamefont {Christensen}, \citenamefont {Cardona}, \citenamefont
  {Chantis}, \citenamefont {van Schilfgaarde},\ and\ \citenamefont
  {Kotani}}]{svane11}%
  \BibitemOpen
  \bibfield  {author} {\bibinfo {author} {\bibfnamefont {A.}~\bibnamefont
  {Svane}}, \bibinfo {author} {\bibfnamefont {N.~E.}\ \bibnamefont
  {Christensen}}, \bibinfo {author} {\bibfnamefont {M.}~\bibnamefont
  {Cardona}}, \bibinfo {author} {\bibfnamefont {A.~N.}\ \bibnamefont
  {Chantis}}, \bibinfo {author} {\bibfnamefont {M.}~\bibnamefont {van
  Schilfgaarde}}, \ and\ \bibinfo {author} {\bibfnamefont {T.}~\bibnamefont
  {Kotani}},\ }\href@noop {} {\bibfield  {journal} {\bibinfo  {journal} {Phys.
  Rev. B}\ }\textbf {\bibinfo {volume} {84}},\ \bibinfo {pages} {205205}
  (\bibinfo {year} {2011})}\BibitemShut {NoStop}%
\bibitem [{\citenamefont {Faleev}\ \emph {et~al.}(2004)\citenamefont {Faleev},
  \citenamefont {van Schilfgaarde},\ and\ \citenamefont {Kotani}}]{faleev04}%
  \BibitemOpen
  \bibfield  {author} {\bibinfo {author} {\bibfnamefont {S.~V.}\ \bibnamefont
  {Faleev}}, \bibinfo {author} {\bibfnamefont {M.}~\bibnamefont {van
  Schilfgaarde}}, \ and\ \bibinfo {author} {\bibfnamefont {T.}~\bibnamefont
  {Kotani}},\ }\href@noop {} {\bibfield  {journal} {\bibinfo  {journal} {Phys.
  Rev. Lett.}\ }\textbf {\bibinfo {volume} {93}},\ \bibinfo {pages} {126406}
  (\bibinfo {year} {2004})}\BibitemShut {NoStop}%
\bibitem [{\citenamefont {Chantis}\ \emph {et~al.}(2006)\citenamefont
  {Chantis}, \citenamefont {van Schilfgaarde},\ and\ \citenamefont
  {Kotani}}]{chantis06}%
  \BibitemOpen
  \bibfield  {author} {\bibinfo {author} {\bibfnamefont {A.~N.}\ \bibnamefont
  {Chantis}}, \bibinfo {author} {\bibfnamefont {M.}~\bibnamefont {van
  Schilfgaarde}}, \ and\ \bibinfo {author} {\bibfnamefont {T.}~\bibnamefont
  {Kotani}},\ }\href@noop {} {\bibfield  {journal} {\bibinfo  {journal} {Phys.
  Rev. Lett.}\ }\textbf {\bibinfo {volume} {96}},\ \bibinfo {pages} {086405}
  (\bibinfo {year} {2006})}\BibitemShut {NoStop}%
\bibitem [{\citenamefont {Chantis}\ \emph {et~al.}(2007)\citenamefont
  {Chantis}, \citenamefont {van Schilfgaarde},\ and\ \citenamefont
  {Kotani}}]{chantis07}%
  \BibitemOpen
  \bibfield  {author} {\bibinfo {author} {\bibfnamefont {A.~N.}\ \bibnamefont
  {Chantis}}, \bibinfo {author} {\bibfnamefont {M.}~\bibnamefont {van
  Schilfgaarde}}, \ and\ \bibinfo {author} {\bibfnamefont {T.}~\bibnamefont
  {Kotani}},\ }\href@noop {} {\bibfield  {journal} {\bibinfo  {journal} {Phys.
  Rev. B}\ }\textbf {\bibinfo {volume} {76}},\ \bibinfo {pages} {165126}
  (\bibinfo {year} {2007})}\BibitemShut {NoStop}%
\bibitem [{Note1()}]{Note1}%
  \BibitemOpen
  \bibinfo {note} {The most strongly correlated phase of Pu is $\delta $ Pu,
  which has a temperature between about 600 and 700 K. As noted in Ref.
  \protect \rev@citealpnum {boivineau01}, ``the delocalization process of the
  $5f$ electrons,'' i.e., the correlation effects, continues to produce
  anomalies as high as 2000 K in temperature in many of these properties, well
  into the high-temperature liquid phase of Pu. Also, see Ref. \protect
  \rev@citealpnum {boivineau09}.}\BibitemShut {Stop}%
\bibitem [{\citenamefont {Abrikosov}\ \emph {et~al.}(1963)\citenamefont
  {Abrikosov}, \citenamefont {Gorkov},\ and\ \citenamefont
  {Dzyaloshinski}}]{abrikosov63}%
  \BibitemOpen
  \bibfield  {author} {\bibinfo {author} {\bibfnamefont {A.~A.}\ \bibnamefont
  {Abrikosov}}, \bibinfo {author} {\bibfnamefont {L.~P.}\ \bibnamefont
  {Gorkov}}, \ and\ \bibinfo {author} {\bibfnamefont {I.~E.}\ \bibnamefont
  {Dzyaloshinski}},\ }\href@noop {} {\emph {\bibinfo {title} {Methods of
  Quantum Field Theory in Statistical Physics}}}\ (\bibinfo  {publisher}
  {Prentice-Hall},\ \bibinfo {address} {Englewood Cliffs, N.J.},\ \bibinfo
  {year} {1963})\BibitemShut {NoStop}%
\bibitem [{\citenamefont {Nozieres}(1964)}]{nozieres64}%
  \BibitemOpen
  \bibfield  {author} {\bibinfo {author} {\bibfnamefont {P.}~\bibnamefont
  {Nozieres}},\ }\href@noop {} {\emph {\bibinfo {title} {Theory of Interacting
  Fermi Systems}}}\ (\bibinfo  {publisher} {W. A. Benjamin},\ \bibinfo
  {address} {New York},\ \bibinfo {year} {1964})\BibitemShut {NoStop}%
\bibitem [{\citenamefont {Hewson}(1993)}]{hewson93}%
  \BibitemOpen
  \bibfield  {author} {\bibinfo {author} {\bibfnamefont {A.~C.}\ \bibnamefont
  {Hewson}},\ }\href@noop {} {\emph {\bibinfo {title} {The Kondo Problem to
  Heavy Fermions}}}\ (\bibinfo  {publisher} {Cambridge University Press},\
  \bibinfo {address} {Cambridge},\ \bibinfo {year} {1993})\BibitemShut
  {NoStop}%
\bibitem [{\citenamefont {Hubbard}(1963)}]{hubbard63}%
  \BibitemOpen
  \bibfield  {author} {\bibinfo {author} {\bibfnamefont {J.}~\bibnamefont
  {Hubbard}},\ }\href@noop {} {\bibfield  {journal} {\bibinfo  {journal} {Proc.
  Roy. Soc}\ }\textbf {\bibinfo {volume} {A276}},\ \bibinfo {pages} {38}
  (\bibinfo {year} {1963})}\BibitemShut {NoStop}%
\bibitem [{\citenamefont {Byczuk}\ \emph {et~al.}(2007)\citenamefont {Byczuk},
  \citenamefont {Kollar}, \citenamefont {Held}, \citenamefont {Yang},
  \citenamefont {Nekrasov}, \citenamefont {Pruschke},\ and\ \citenamefont
  {Vollhardt}}]{byczuk07}%
  \BibitemOpen
  \bibfield  {author} {\bibinfo {author} {\bibfnamefont {K.}~\bibnamefont
  {Byczuk}}, \bibinfo {author} {\bibfnamefont {M.}~\bibnamefont {Kollar}},
  \bibinfo {author} {\bibfnamefont {K.}~\bibnamefont {Held}}, \bibinfo {author}
  {\bibfnamefont {Y.-F.}\ \bibnamefont {Yang}}, \bibinfo {author}
  {\bibfnamefont {I.~A.}\ \bibnamefont {Nekrasov}}, \bibinfo {author}
  {\bibfnamefont {T.}~\bibnamefont {Pruschke}}, \ and\ \bibinfo {author}
  {\bibfnamefont {D.}~\bibnamefont {Vollhardt}},\ }\href@noop {} {\bibfield
  {journal} {\bibinfo  {journal} {Nat. Phys.}\ }\textbf {\bibinfo {volume}
  {3}},\ \bibinfo {pages} {168} (\bibinfo {year} {2007})}\BibitemShut {NoStop}%
\bibitem [{\citenamefont {Wick}(1967)}]{wick67}%
  \BibitemOpen
  \bibfield  {author} {\bibinfo {author} {\bibfnamefont {O.~J.}\ \bibnamefont
  {Wick}},\ }\href@noop {} {\emph {\bibinfo {title} {Plutonium Handbook: A
  Guide to the Technology}}}\ (\bibinfo  {publisher} {Gordon and Breach},\
  \bibinfo {address} {New York},\ \bibinfo {year} {1967})\BibitemShut {NoStop}%
\bibitem [{\citenamefont {Hecker}(2001)}]{hecker01}%
  \BibitemOpen
  \bibfield  {author} {\bibinfo {author} {\bibfnamefont {S.~S.}\ \bibnamefont
  {Hecker}},\ }\href@noop {} {\bibfield  {journal} {\bibinfo  {journal} {MRS
  Bull.}\ }\textbf {\bibinfo {volume} {26}},\ \bibinfo {pages} {672} (\bibinfo
  {year} {2001})}\BibitemShut {NoStop}%
\bibitem [{\citenamefont {Hecker}(2004)}]{hecker04a}%
  \BibitemOpen
  \bibfield  {author} {\bibinfo {author} {\bibfnamefont {S.~S.}\ \bibnamefont
  {Hecker}},\ }\href@noop {} {\bibfield  {journal} {\bibinfo  {journal} {Met.
  Mat. Trans. A}\ }\textbf {\bibinfo {volume} {35A}},\ \bibinfo {pages} {2207}
  (\bibinfo {year} {2004})}\BibitemShut {NoStop}%
\bibitem [{\citenamefont {Hecker}\ \emph {et~al.}(2004)\citenamefont {Hecker},
  \citenamefont {Harbur},\ and\ \citenamefont {Zocco}}]{hecker04b}%
  \BibitemOpen
  \bibfield  {author} {\bibinfo {author} {\bibfnamefont {S.~S.}\ \bibnamefont
  {Hecker}}, \bibinfo {author} {\bibfnamefont {D.~R.}\ \bibnamefont {Harbur}},
  \ and\ \bibinfo {author} {\bibfnamefont {T.~G.}\ \bibnamefont {Zocco}},\
  }\href@noop {} {\bibfield  {journal} {\bibinfo  {journal} {Prog. Mat. Sci.}\
  }\textbf {\bibinfo {volume} {49}},\ \bibinfo {pages} {429} (\bibinfo {year}
  {2004})}\BibitemShut {NoStop}%
\bibitem [{\citenamefont {Albers}(2001)}]{albers01}%
  \BibitemOpen
  \bibfield  {author} {\bibinfo {author} {\bibfnamefont {R.~C.}\ \bibnamefont
  {Albers}},\ }\href@noop {} {\bibfield  {journal} {\bibinfo  {journal} {Nature
  (London)}\ }\textbf {\bibinfo {volume} {410}},\ \bibinfo {pages} {759}
  (\bibinfo {year} {2001})}\BibitemShut {NoStop}%
\bibitem [{Note2()}]{Note2}%
  \BibitemOpen
  \bibinfo {note} {We have not included spin-orbit effects, which can be safely
  ignored for the purposes of this paper. The Pu $f$ DOS splits into a pair of
  clearly separated $j=5/2$ and $7/2$ peaks. To include spin-orbit, we would
  need to calculate the bandwidth of each peak separately and use that
  corresponding to $j=5/2$. By ignoring spin-orbit coupling, we are saved from
  this additional trouble, which is not expected to change the effective $f$
  bandwidths. Recent spin-orbit $GW$ calculations have been calculated in Pu
  (Ref.~\protect \rev@citealpnum {kutepov12}). However these have been done in
  the fully self-consistent $GW$ method, which usually is a poor approximation
  in solids due to an incorrect treatment of plasmon effects. Since the DOS in
  this paper includes broadening effects due to the imaginary part of the
  self-energy in all of the different approximation that were used, it is also
  unclear how bandwidth narrowing would separately be affected by spin-orbit
  effects.}\BibitemShut {Stop}%
\bibitem [{\citenamefont {Bouchet}\ \emph {et~al.}(2004)\citenamefont
  {Bouchet}, \citenamefont {Albers}, \citenamefont {Jones},\ and\ \citenamefont
  {Jomard}}]{psalpha}%
  \BibitemOpen
  \bibfield  {author} {\bibinfo {author} {\bibfnamefont {J.}~\bibnamefont
  {Bouchet}}, \bibinfo {author} {\bibfnamefont {R.~C.}\ \bibnamefont {Albers}},
  \bibinfo {author} {\bibfnamefont {M.~D.}\ \bibnamefont {Jones}}, \ and\
  \bibinfo {author} {\bibfnamefont {G.}~\bibnamefont {Jomard}},\ }\href@noop {}
  {\bibfield  {journal} {\bibinfo  {journal} {Phys. Rev. Lett.}\ }\textbf
  {\bibinfo {volume} {92}},\ \bibinfo {pages} {095503} (\bibinfo {year}
  {2004})}\BibitemShut {NoStop}%
\bibitem [{Note3()}]{Note3}%
  \BibitemOpen
  \bibinfo {note} {Choosing instead for example $D_{f}(E)$ as the weight
  function does not serve the purpose of emphasizing the central over the
  ``hybridization wings'' in the $f$ DOS.}\BibitemShut {Stop}%
\bibitem [{Note4()}]{Note4}%
  \BibitemOpen
  \bibinfo {note} {For the atomic volumes we have ignored any thermal volume
  expansion. Each phase is represented by a volume corresponding to a fixed
  temperature within that phase. We have used the original data of Zachariasen
  and Ellinger (Refs.~\protect \rev@citealpnum
  {zachariasen63b,zachariasen63a,zachariasen55,ellinger56}) corresponding to
  the volumes at the temperatures 21, 190, 235, 320, 477, and 490 $^\circ $C,
  for the $\alpha $, $\beta $, $\gamma $, $\delta $, $\delta '$, and $\epsilon
  $ phases, respectively.}\BibitemShut {Stop}%
\bibitem [{Note5()}]{Note5}%
  \BibitemOpen
  \bibinfo {note} {For the volumes of the different phases of Pu, we have
  followed the same method used to generate Fig.~\ref {fig:pu_bwred_V}. We have
  also used the same volumes of the different phases for the sound velocity and
  resistivity needed to determine the correlation strength from the $GW$
  calculations plotted in Fig.~\ref {fig:pu_bwred_V}. Note that, since we have
  not directly calculated the value of ${\protect \textrm w_{rel}}$ for the
  $\beta $ phase, we instead used the availability of the bandwidth reduction
  of Eq. (\ref {eqn:w_rel_fit}) together with the calculated LDA bandwidth for
  the correct crystal structure of $\beta $ Pu to determine ${w_{rel}}(\beta )$
  = 0.55.}\BibitemShut {Stop}%
\bibitem [{\citenamefont {Boivineau}(2001)}]{boivineau01}%
  \BibitemOpen
  \bibfield  {author} {\bibinfo {author} {\bibfnamefont {M.}~\bibnamefont
  {Boivineau}},\ }\href@noop {} {\bibfield  {journal} {\bibinfo  {journal} {J.
  Nuc. Mater.}\ }\textbf {\bibinfo {volume} {297}},\ \bibinfo {pages} {97}
  (\bibinfo {year} {2001})}\BibitemShut {NoStop}%
\bibitem [{\citenamefont {Boivineau}(2009)}]{boivineau09}%
  \BibitemOpen
  \bibfield  {author} {\bibinfo {author} {\bibfnamefont {M.}~\bibnamefont
  {Boivineau}},\ }\href@noop {} {\bibfield  {journal} {\bibinfo  {journal} {J.
  Nuc. Mater.}\ }\textbf {\bibinfo {volume} {392}},\ \bibinfo {pages} {568}
  (\bibinfo {year} {2009})}\BibitemShut {NoStop}%
\bibitem [{\citenamefont {Zachariasen}\ and\ \citenamefont
  {Ellinger}(1963{\natexlab{a}})}]{zachariasen63b}%
  \BibitemOpen
  \bibfield  {author} {\bibinfo {author} {\bibfnamefont {W.~H.}\ \bibnamefont
  {Zachariasen}}\ and\ \bibinfo {author} {\bibfnamefont {F.~H.}\ \bibnamefont
  {Ellinger}},\ }\href@noop {} {\bibfield  {journal} {\bibinfo  {journal} {Acta
  Cryst.}\ }\textbf {\bibinfo {volume} {16}},\ \bibinfo {pages} {777} (\bibinfo
  {year} {1963}{\natexlab{a}})}\BibitemShut {NoStop}%
\bibitem [{\citenamefont {Zachariasen}\ and\ \citenamefont
  {Ellinger}(1963{\natexlab{b}})}]{zachariasen63a}%
  \BibitemOpen
  \bibfield  {author} {\bibinfo {author} {\bibfnamefont {W.~H.}\ \bibnamefont
  {Zachariasen}}\ and\ \bibinfo {author} {\bibfnamefont {F.~H.}\ \bibnamefont
  {Ellinger}},\ }\href@noop {} {\bibfield  {journal} {\bibinfo  {journal} {Acta
  Cryst.}\ }\textbf {\bibinfo {volume} {16}},\ \bibinfo {pages} {395} (\bibinfo
  {year} {1963}{\natexlab{b}})}\BibitemShut {NoStop}%
\bibitem [{\citenamefont {Zachariasen}\ and\ \citenamefont
  {Ellinger}(1955)}]{zachariasen55}%
  \BibitemOpen
  \bibfield  {author} {\bibinfo {author} {\bibfnamefont {W.~H.}\ \bibnamefont
  {Zachariasen}}\ and\ \bibinfo {author} {\bibfnamefont {F.~H.}\ \bibnamefont
  {Ellinger}},\ }\href@noop {} {\bibfield  {journal} {\bibinfo  {journal} {Acta
  Cryst.}\ }\textbf {\bibinfo {volume} {8}},\ \bibinfo {pages} {431} (\bibinfo
  {year} {1955})}\BibitemShut {NoStop}%
\bibitem [{\citenamefont {Ellinger}(1956)}]{ellinger56}%
  \BibitemOpen
  \bibfield  {author} {\bibinfo {author} {\bibfnamefont {F.~H.}\ \bibnamefont
  {Ellinger}},\ }\href@noop {} {\bibfield  {journal} {\bibinfo  {journal} {J.
  of Metals}\ }\textbf {\bibinfo {volume} {8}},\ \bibinfo {pages} {1256}
  (\bibinfo {year} {1956})}\BibitemShut {NoStop}%
\end{thebibliography}

\end{document}